\documentclass[conference]{article}
\usepackage{cite}
\usepackage{amsmath,amssymb,amsfonts}
\usepackage{algorithmic}
\usepackage{graphicx}
\usepackage{textcomp}
\usepackage{xcolor}
\usepackage{multirow}
\usepackage{multicol}
\usepackage{subcaption}
\usepackage{hyperref}
\usepackage{todonotes}

\def\BibTeX{{\rm B\kern-.05em{\sc i\kern-.025em b}\kern-.08em
    T\kern-.1667em\lower.7ex\hbox{E}\kern-.125emX}}
\begin{document}

\title{Robust and accurate fine-grain power models for embedded systems with no on-chip PMU}

\author{Kris Nikov, Marcos Mart\'{i}nez, Simon Wegener, Jose Nunez-Yanez,\\ Zbigniew Chamski, Kyriakos Georgiou and Kerstin Eder%
\thanks{K.\ Nikov, J.\ Nunez-Yanez, Z.\ Chamski, K.\ Georgiou and K.\ Eder are with the University of Bristol, UK. 
(email: kris.nikov@bristol.ac.uk, j.l.nunez-yanez@bristol.ac.uk, zbigniew.chamski@bristol.ac.uk, kyriakos.georgiou@bristol.ac.uk, kerstin.eder@bristol.ac.uk)}
\thanks{M.\ Mart\'{i}nez is with Thales Alenia Space, Madrid, Spain. 
(email: marcos.martinezalejandro@thalesaleniaspace.com)}
\thanks{S.\ Wegener is with AbsInt Angewandte Informatik, Saarbrücken, Germany. 
(email: swegener@absint.com)}
}%

\maketitle

\begin{abstract}
This paper presents a novel approach to event-based power modelling for embedded platforms that do not have a Performance Monitoring Unit (PMU). The  method involves complementing the target hardware platform, where the physical power data is measured, with another platform on which the CPU performance data, that is needed for model generation, can be collected. 
The methodology is used to generate accurate fine-grain power models for the Gaisler GR712RC dual-core LEON3 fault-tolerant SPARC processor with on-board power sensors and no PMU. A Kintex UltraScale FPGA is used as the support platform to obtain the required CPU performance data, by running a soft-core representation of the dual-core LEON3 as on the GR712RC but with a PMU implementation.
Both platforms execute the same benchmark set and data collection is synchronised using per-sample timestamps so that the power sensor data from the GR712RC board can be matched to the PMU data from the FPGA. The synchronised samples are then processed by the Robust Energy and Power Predictor Selection (REPPS) software in order to generate power models. 
The models achieve less than 2\% power estimation error when validated on an industrial use-case and can follow program phases, which makes them suitable for runtime power profiling during development.
\end{abstract}


\section{Introduction}
\label{sec:introduction}

Power analysis enables hardware designers and software developers to optimise the energy consumption of embedded systems. 
Robust and accurate power models are fundamental in this context, with hardware event-based power modelling being a widely used technique both for CPU as well as full system modelling~\cite{rodrigues2013study, nunez2013enabling,rethinagiri2014system,Walker2017,seewald2021coarse}.
Rodrigues et al.~\cite{rodrigues2013study} present a systematic review of common Performance Monitoring Unit (PMU) events, also termed Performance Monitoring Counters (PMCs), in modern microprocessors and show their effectiveness in characterising and modelling dynamic power consumption. 
The challenge is how to develop accurate power models for systems without an on-chip PMU. This paper introduces an innovative, dual-platform approach for power modelling of such platforms, and includes full model validation against physical power measurements. 

In the space industry, devices that operate under tight resource constraints often remain deployed for years, relying only on remote maintenance.
Continuous development over the life-cycle of such systems can be achieved via dynamic over-the-air software and firmware  updates~\cite{jurkovic2014remote,lounas2019formal}.
In the case of satellite communications, energy efficiency is a critical requirement, with development focusing on processors such as the LEON3 microprocessor. Our dual-platform methodology was used to characterise the power consumption of the LEON3 on the GR712RC development platform~\cite{GR712RCDevBoard}, which does not have an on-chip PMU. The omission of the PMU is typical of deeply embedded devices, where any hardware that represents a power or area overhead during deployment is removed before fabrication. 
The models were deployed during remote software development to enable early power analysis and optimisation, with the aim to ensure that any over-the-air updates meet the energy and power constraints before they are applied and without the need for direct access to the platform. The techniques described in this paper can be applied to other open embedded hardware platforms with no PMU. 

Existing research towards energy models for the LEON3 processor include instruction-level energy models for a custom LEON3 design~\cite{penolazzi2009energy}. The program execution and energy consumption data is generated using an RTL gate-level simulator targeting a 90nm implementation at 400Mhz CPU core frequency. The models are validated using a cycle accurate instruction set simulator, achieving a worst-case estimation error of $\pm 12\%$ when compared to the gate-level design simulation. 
Another power estimation approach for the LEON3~\cite{ImplementingLEON3Statistics} achieved an average error between 1.5\% and 2.1\%. The authors used FPGA emulation for a custom design at 25Mhz operating frequency to obtain hardware counter measurements and use an external gate-level analysis tool for power estimates.
Both approaches achieve low model errors compared to simulation-based power estimation, but lack validation against hardware measurements, which is essential to gain full confidence in the accuracy of the models. 
A similar dual-platform approach based on real hardware measurements together with event data collected from a cycle-accurate instruction set emulator has been used to generate PMC-based energy models for the Arm Cortex-M0 processor~\cite{georgiou2021comprehensive}. However, these models use samples with coarser granularity to inform static energy consumption analysis and compile-time optimisations; they are not suitable for runtime power profiling during development. 

This paper offers the following scientific contributions:
\begin{enumerate} 
	\item A dual-platform approach to collect PMCs from an FPGA soft-core implementation and to synchronize these with direct power measurements from a physical board using per-sample timestamps to enable fine-grain PMC-based power modelling for hardware platforms with no on-chip PMU. 
    \item A detailed power model for the LEON3 processor that has undergone comprehensive validation against hardware power measurements.
	\item Portable, modular and open-source model generation software, named Robust Energy and Power Predictor Selection (REPPS)~\cite{REPPS}, that implements several search algorithms along with k-fold cross validation in order to identify the optimal selection of PMCs for the model. 
\end{enumerate}

\section{Power Modelling Methodology}
\begin{figure}[thb!]
  	\centering
    \includegraphics[width=1\linewidth]{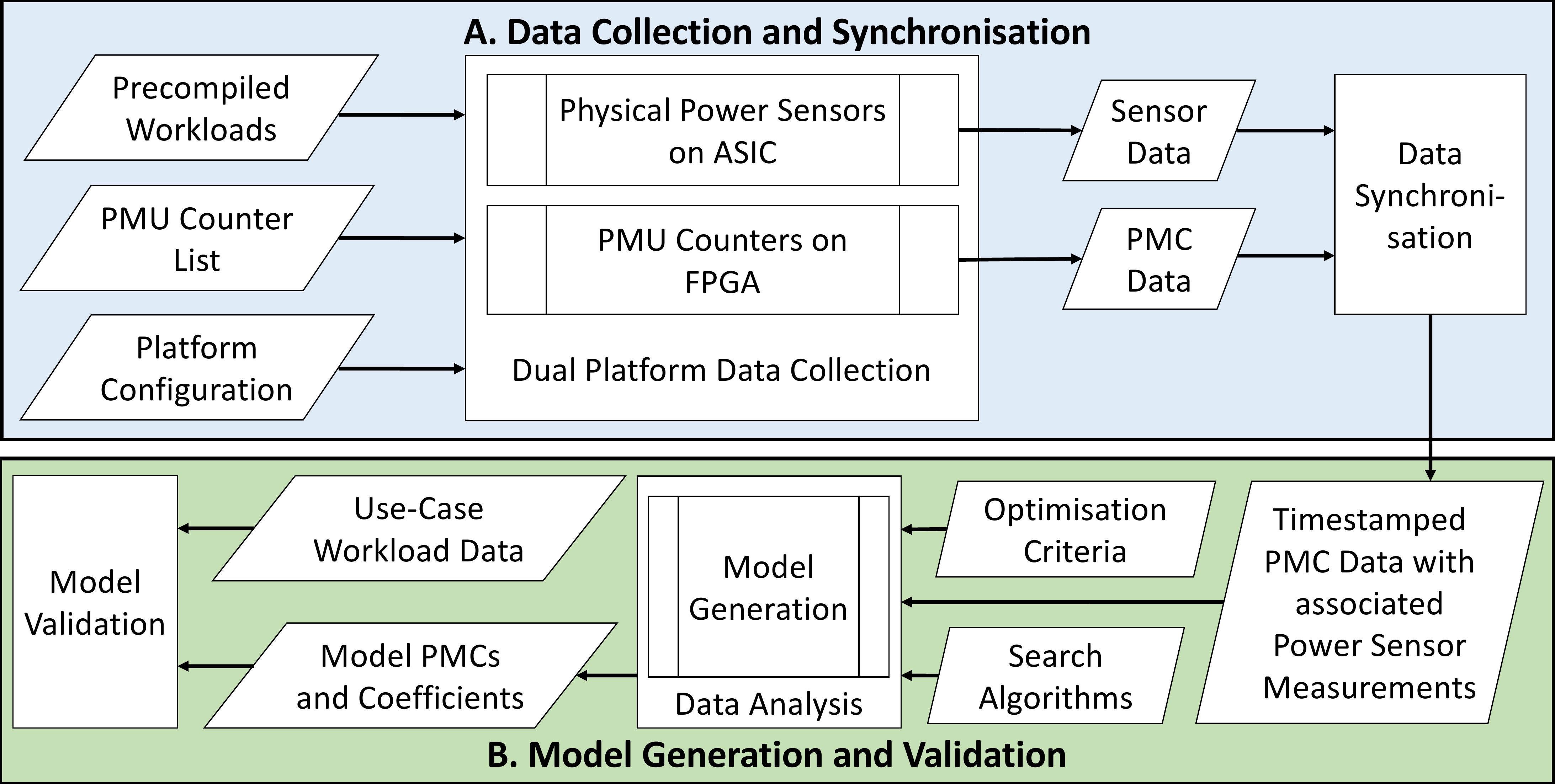}%
  	\caption{Power modelling methodology stages.}
  	\vspace{-0.8em}
    \label{fig:energy:leon3:methodology}
\end{figure}

The power modelling methodology is comprised of two stages: {\em A.\ Data Collection and Synchronisation} and {\em B.\ Model Generation and Validation} as shown in more detail in Figure~\ref{fig:energy:leon3:methodology}.

\subsection{Data Collection and Synchronisation}
\label{subsec:energy:leon3:data_collection}

\subsubsection{Platform Configuration}
\label{subsubsec:leon3:platform}
First, both the ASIC and FPGA hardware platforms need to be set up to be used in tandem to collect the data for model generation. 
The target ASIC is the GR712RC evaluation board, used predominantly in the space industry. The specific platform is ideal for evaluating the model generation methodology, since the on-chip LEON3 CPU RTL design is available under the GNU GPL license, allowing free and unlimited use for research and education. The LEON3 CPU on the platform features a custom dual-core implementation of the 32-bit SPARC V8 ISA~\cite{gaisler_gr712rc_cpu}
, equipped with fault and radiation resistant technologies, making it suitable for outer space operations. This processor implementation does not include a PMU, but the ASIC offers on-board power sensors.
The PMU IP available for the LEON3 processors is the LEON3 Statistics Unit, L3STAT~\cite{gaisler_tools}
, which offers a configurable number of (up to 64) 32-bit counters that can count events in the processor core or the AHB bus of the LEON3.
This processor design is synthesised together with the L3STAT unit on a Kintex UltraScale KU060 FPGA ~\cite{ku060overview} 
for the collection of the event counts, while power measurements are obtained from the GR712RC development board.  

For this dual-platform approach to work, it is critical to ensure that the synthesised processor matches the behaviour of the hardware implementation on the GR712RC that is to be modelled. The main features of the LEON3 on both platforms are a 16KiB (4x4kB) multi-way instruction cache, 16KiB (4x4kB) multi-way data cache and a 80MHz frequency clock. There are two significant differences between the processor on the development board and its FPGA synthesised version. The LEON3 on the GR712RC has a high-performance double-precision IEEE-754 floating-point unit, which is not open source, i.e.\ not included in the RTL distribution. To be able to run the exact program compilation on both platforms, the hardware FPU is disabled through compilation options and a software library is used instead to compute floating point operations. 
Also, the memory read and write timing is different, which does not allow the FPGA to run at the same speed as the development board, 80MHz. Thus, it is necessary to extend the timing of memory operations to five clock cycles from the regular two clock cycles on the GR712RC board. The FPGA implementation requires three wait states to be able to run programs at 80MHz and execution time is the same as on the ASIC platform, with a negligible delta of a few microseconds. The GRMON ~\cite{gaisler_tools}
 debugger was used to control and monitor the LEON3 CPUs in both setups.

\subsubsection{PMU Counter List}
\label{subsubsec:leon3:pmu_events}

The modelling methodology requires a list of supported counters that can be collected using the soft-core CPU implementation on the supporting FPGA platform. The L3STAT unit for the LEON3 is used to monitor the 17 CPU-specific PMCs shown in Table~\ref{tab:leon3:all_pmu_events}, which is the complete list of available non-zero counters during workload execution. $C_0$ is used to synchronize the PMC data to the power sensor data and $C_1 - C_{16}$ are used in model generation. 
\begin{table}[hbt!]
\centering
\resizebox{1\linewidth}{!}{%
\begin{tabular}{|c|l||c|l||c|l|}
\hline
\#      & Counter & \#       & Counter  & \#       & Counter \\ \hline \hline
$C_{0}$ & TIME  & $C_{6}$  & AINST    & $C_{12}$ & CALL    \\ \hline
$C_{1}$ & ICMISS  & $C_{7}$  & IINST    & $C_{13}$ & TYPE2   \\ \hline
$C_{2}$ & ICHOLD  & $C_{8}$  & BPMISS   & $C_{14}$ & LDST    \\ \hline
$C_{3}$ & DCMISS  & $C_{9}$  & AHBUTIL  & $C_{15}$ & LOAD    \\ \hline
$C_{4}$ & DCHOLD  & $C_{10}$ & AHBTUTIL & $C_{16}$ & STORE   \\ \hline
$C_{5}$ & WBHOLD  & $C_{11}$ & BRANCH   &          &         \\ \hline
\end{tabular}
}
\caption{PMCs available for the LEON3 power model.}
  	\vspace{-1em}
\label{tab:leon3:all_pmu_events}
\end{table} 
\subsubsection{Precompiled Workloads}
\label{subsubsec:leon3:benchmarks}
In order to obtain the necessary data for successful platform power modelling, the workloads used to exercise the target CPU need to be carefully selected. 

BEEBS~\cite{BEEBSa} is an open-source benchmark suite designed for performance and energy consumption analysis of embedded architectures. It includes several subsets of workloads, representing a wide variety of embedded applications. 
This diversity makes BEEBS an excellent training set, ensuring that the model is robust, flexible and not over-fitted to a specific application type.
The set of benchmarks used for training consists of the 50 distinct workloads from BEEBS, that were successfully compiled using the Gaisler RTEMS compiler ~\cite{rtems}
 and executed on the two platforms. The benchmarks were executed four times each in order to obtain statistically robust measurements. The training data has over 288000 sample points. The average execution time variation between the four runs was 0.34\% for both CPU and FPGA implementations. 

The models were evaluated on a proprietary computer vision algorithm used in space satellite imaging. There are four different compiled versions of this algorithm, obtained using two different compilers and levels of optimisation. The four binaries are executed three times each resulting in a test set of over 23000 sample points. A detailed list of the train and test benchmark sets, as well as individual workload execution times is available in the project code repository~\cite{REPPS}.

\subsubsection{Data Synchronisation}
\label{subsubsec:leon3:data_sync}
The most critical part of the dual-platform setup is to ensure that the power sensor data from the ASIC corresponds to the correct PMC data from the FPGA. The cross-platform synchronisation methodology for the target platform consists of the following steps:
\begin{enumerate}
	\item Configure the FPGA with the LEON3 and L3STAT.
	\item Initialise the processor on the FPGA with three wait states on the memory access using GRMON.
	\item Set up the L3STAT for polling the available PMCs as fast as possible, collecting around 95 samples per second.
	\item Load and run the benchmarks on the FPGA and store the PMC data.
	\item Program the GR712RC power sensor using the CPU cycle (\textit{TIME} counter) data from the FPGA so that power can be sampled on the ASIC at the exact same times.

	\item Initialise the processor on the GR712RC with three wait states on the memory access using GRMON.
	\item Run the benchmark on the ASIC and store the power measurements.
	\item For the same value of the \textit{TIME} counter, associate the sensor data with the corresponding PMC data.
\end{enumerate}

\subsection{Model Generation and Validation}
\label{subsec:leon3:model_generation}
The second stage of the methodology uses the data generated by the first stage. 
The model generation software is an extension of~\cite{nikov2020intra}, adapted to the data from the dual-platform setup.
All code is open-source and available online~\cite{REPPS}. 
\subsubsection{Optimisation Criteria and Search Algorithms}
\label{subsubsec:leon3:autosearch}
The methodology uses two search algorithms to find the optimal power model from the collected PMC data. The metric to optimise is the Mean Absolute Percentage Error (MAPE).
The first algorithm uses a \textit{bottom-up} strategy. It traverses the list of available PMCs and adds the best PMC, according to the chosen optimisation criteria (minimising the MAPE), to the model after every completed iteration of the list.
The methodology provides the ability to choose an initial set of events to start from, as well as the maximum number of events to include in the computed model.
The second algorithm is a \textit{top-down} approach, with starts from a user-selected list of PMCs and removes, at each iteration, the PMC which reduces the model MAPE the most.

Both search algorithms perform k-fold-cross-validation~\cite{fushiki2011estimation} on the training set each time a new candidate PMC is analysed for inclusion into or removal from the model.
At each search step, the average model MAPE across all the folds is used as the performance metric to optimise. The final set of model coefficients is calculated on the full set of training samples.
\subsubsection{Data Analysis}
\label{subsubsec:leon3:data_analysis}
Ordinary Least Squares (OLS)~\cite{kutner2005applied} \emph{linear regression} is used to generate a power model expressed by $P = {\alpha} + {\beta}_1 \times C_1 + \ldots + {\beta}_n \times C_n$, where the regressor weights ($\beta_x$) are obtained for each activity ($C_x$), i.e.\ PMC, and the residual ($\alpha$) represents the idle power consumption. 
The estimated power dissipation ($P$) can then be calculated based on the PMC values for a given program and its inputs.

\subsubsection{Model Validation}
\label{subsubsec:leon3:model_validation}
The accuracy of the model is validated using PMU data from the test set.
The measured power values for the test set are then compared to the power estimations obtained from the model.
The prediction accuracy of the power models can then be assessed using the MAPE.
\section{Experimental Results}\label{s:exp-results}
\begin{figure*}[!htb]
\centering
	\includegraphics[width=1\linewidth]{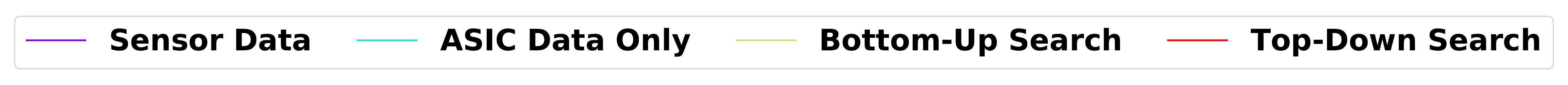}  		
	\begin{subfigure}{1\linewidth}
		\centering
		\includegraphics[width=1\linewidth]{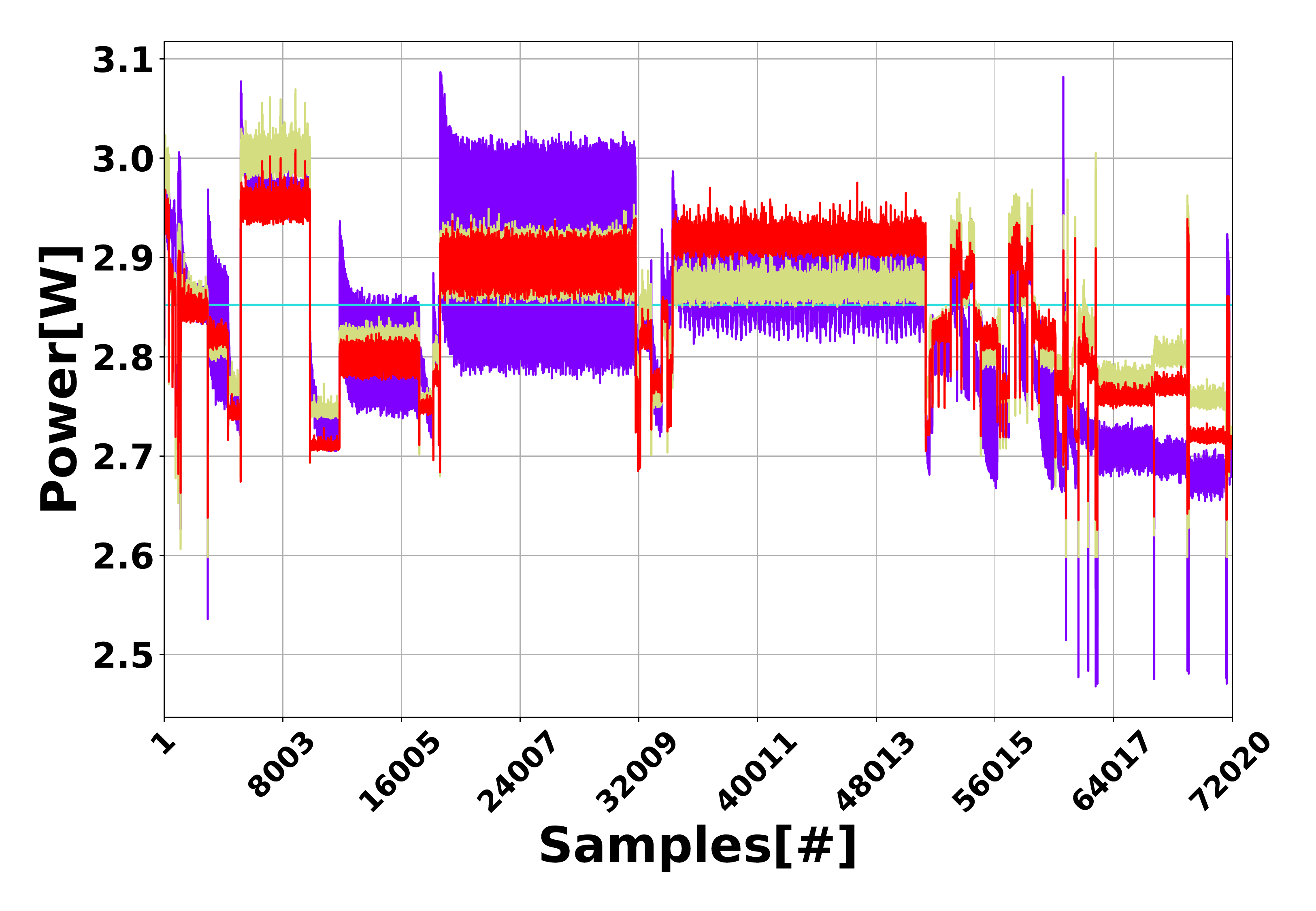}
  		\caption{Train set - BEEBS}
    		\label{fig:leon3:fine_grain_power_models_performance_trainset}
	\end{subfigure}
	\begin{subfigure}{1\linewidth}
		\centering	
		\includegraphics[width=1\linewidth]{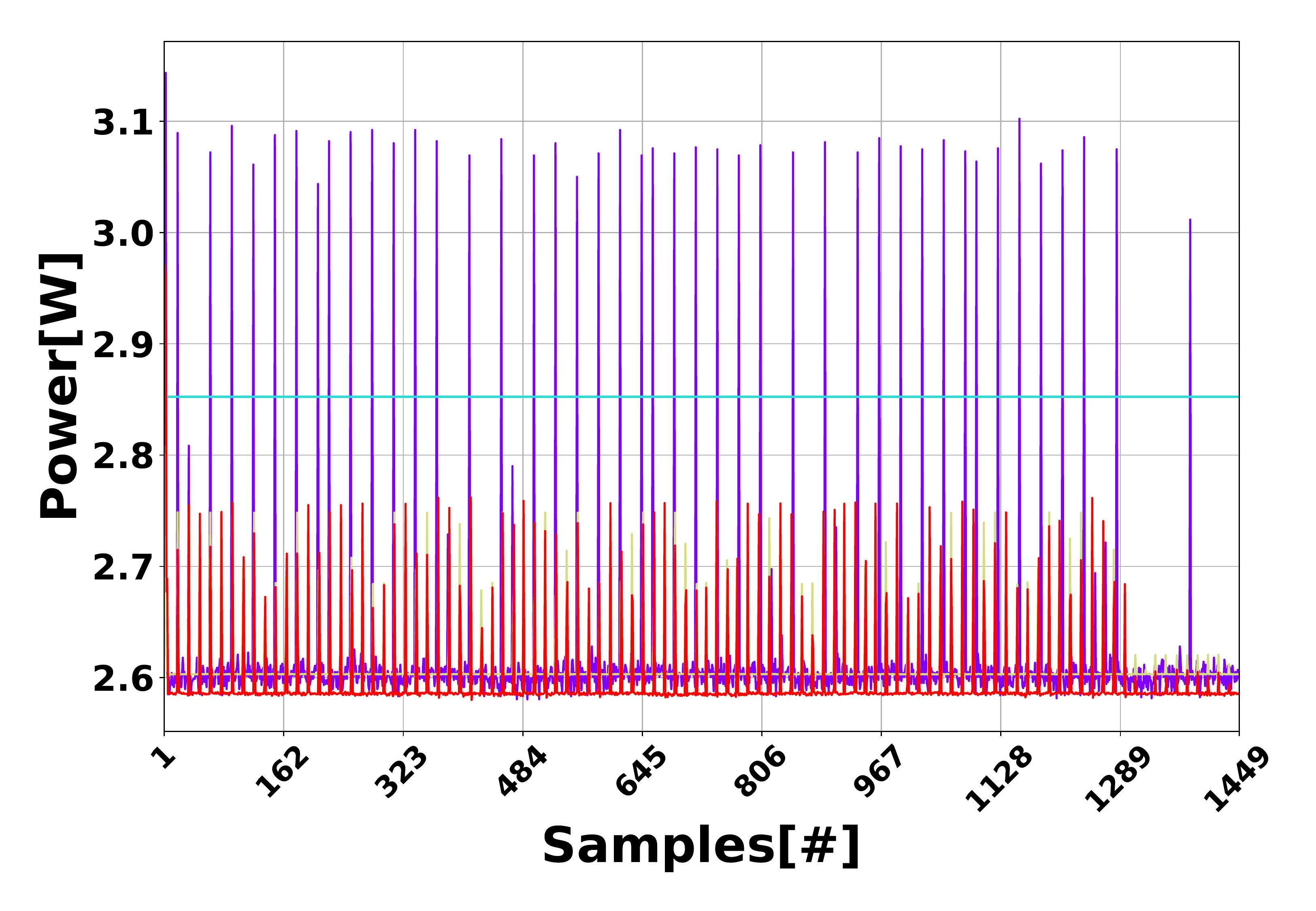}	
		\caption{Test set - use\_case\_opt}    			\label{fig:leon3:fine_grain_power_models_performance_testset}
	\end{subfigure}
\caption{Power model per-sample prediction on the LEON3 platform for the first run of the test and train sets.}  
  	\vspace{-1em}
\label{fig:leon3:fine_grain_power_models_performance}
\end{figure*}

Both \textit{bottom-up} and \textit{top-down} search algorithms are used for model PMC selection. The resulting models are compared to an \textit{ASIC only} model, which solely uses frequency information from the GR712RC on-board sensors to obtain a prediction of the average power consumption. The PMC selection is done using 50-fold cross-validation, which is the maximum number of folds available for the BEEBS training set. Table~\ref{tab:leon3:fine_grain_model_results} contains the model equations as well as the model performance results for the train and test sets. Figure~\ref{fig:leon3:fine_grain_power_models_performance} is a visual representation of the predicted power values of the models against the actual measured data for the first run of the train set and the \texttt{use\_case\_opt} compiled version of the test set using the Gaisler RTEMS compiler with the \texttt{-O3} optimisation flag.
\begin{table}[ht!]
\centering
\resizebox{\linewidth}{!}{%

\begin{tabular}{|l||l||c|c|}
\hline
\multirow{2}{*}{Model Name}                                                                                                                                                                               & \multirow{2}{*}{Power Model Equation}                                                                                                                                                                                                                                                                                                                                                                                                                                                                                                                                                                                                                                                                                                                                                                               & \multicolumn{2}{c|}{MAPE {[}\%{]}} \\ \cline{3-4}
                                                                                                                                                                                                     &                                                                                                                                                                                                                                                                                                                                                                                                                                                                                                                                                                                                                                                                                                                                                                                                           & \begin{tabular}[c]{@{}c@{}}Train\\ (BEEBS)\end{tabular}    & \begin{tabular}[c]{@{}c@{}}Test\\ (use\_case)\end{tabular}    \\ \hline \hline                                                                                                                                                                                  \begin{tabular}[c]{@{}l@{}}Power {[}W{]}\\ ASIC Data\end{tabular}  &   \begin{tabular}[c]{@{}l@{}} $\text{P} = 0.000445617	+0.0356494\times\text{Freq.[MHz]}$\end{tabular}   &	2.56	& 6.73 \\ \hline
\begin{tabular}[c]{@{}l@{}}Power {[}W{]}\\ Bottom-Up\end{tabular}  & \begin{tabular}[c]{@{}l@{}} $\text{P} = 2.59799	+	4.58765\text{e-}06\times{C_{16}}$\end{tabular}     &	1.52	& 1.45 \\ \hline
\begin{tabular}[c]{@{}l@{}}Power {[}W{]}\\ Top-Down\end{tabular}   & \begin{tabular}[c]{@{}l@{}} $\text{P} = 2.61526	-6.64855\text{e-}05\times{C_{1}}	$\\$+2.10177\text{e-}06\times{C_{2}}	-2.30418\text{e-}07\times{C_{3}}	$\\$+1.7569\text{e-}07\times{C_{5}}	-9.78606\text{e-}07\times{C_{6}}	$\\$+8.83961\text{e-}07\times{C_{7}}	+1.0862\text{e-}06\times{C_{8}}	$\\$+3.73317\text{e-}07\times{C_{10}}	-5.25209\text{e-}07\times{C_{11}}	$\\$+1.37306\text{e-}06\times{C_{12}}	+1.51902\text{e-}06\times{C_{14}}	$\\$-1.15743\text{e-}06\times{C_{15}}$\end{tabular}     &	1.14	& 1.72 \\ \hline
\end{tabular}
}
\caption{LEON3 power models obtained from different model generation methods with validation results.}
\label{tab:leon3:fine_grain_model_results}
\end{table}

Both \textit{bottom-up} and \textit{top-down} searches have identified a different set of PMCs for the respective power models. The model from the \textit{bottom-up} search uses a single PMC, whereas the \textit{top-down} model uses 12 PMCs. This highlights the need to use different search algorithms to identify the set of PMCs in order to identify local optima.

Figure~\ref{fig:leon3:fine_grain_power_models_performance_trainset} shows that the PMC-based models match the average power consumption of the individual BEEBS benchmarks during execution, including the direction of the dynamic power peaks. 
The predictions from the PMC-based models follow the program phases as power varies over time, with the baseline provided by the sensor data. These models are suitable for predictive power profiling, whereas the \textit{ASIC only} model is not.
Regarding the test set, the PMC-based models are able to predict the average power consumption and power draw spikes of the program as illustrated in Figure~\ref{fig:leon3:fine_grain_power_models_performance_testset}. 
However, the models underestimate the peak power at the power draw spike points. This is caused both by the limitation in regression-based models, which cannot handle large fluctuations in the modelled data, as well as the limited selection of PMCs used. Nevertheless, the models still recognise the points of power variation, which is why the prediction error is so low.

\section{Conclusion}
This paper proposes and demonstrates a novel dual-platform approach to generating accurate fine-grain PMC-based power models for target platforms with no on-chip PMU, but for which the RTL design is available.
In this approach, the physical power data is obtained from the target hardware platform, and these measurements are then synchronised on a per-sample basis with the performance data collected from a soft-core FPGA implementation instrumented with a PMU. 
The synchronised samples are then processed by the Robust Energy and Power Predictor Selection (REPPS) software in order to generate power models. 
REPPS uses automatic search methods to select the set of PMC events that produce the model with highest estimation accuracy.

This dual-platform approach has been used to generate accurate fine-grain power models for the the Gaisler GR712RC dual-core LEON3 fault-tolerant SPARC processor with on-board power sensors and no PMU. 
The power models for the LEON3 achieve less than 2\% Mean Absolute Percentage Error (MAPE) when validated on an industry-representative image processing algorithm, used in space communications. 
The methodology can be used to characterise similar platforms. It is limited by the availability of a soft-core version with PMU and the number as well as types of PMCs available. 

\section*{Acknowledgment}
This research was supported by the European Union's Horizon 2020 Research and
Innovation Programme under grant agreement No.\ 779882, TeamPlay (Time, Energy and security Analysis for Multi/Many-core heterogeneous PLAtforms).
\bibliographystyle{alpha}
\bibliography{bibliography/bibliography}
\end{document}